\documentclass[%
 aip,
 amsmath,amssymb,
 reprint,%
]{revtex4-1}

\usepackage{graphicx}
\usepackage{dcolumn}
\usepackage{bm}
\usepackage[utf8]{inputenc}
\usepackage[T1]{fontenc}
\usepackage{mathptmx}
\usepackage{etoolbox}
\usepackage{color,soul}
\usepackage{textcomp,mathcomp}
\makeatletter
\def\@email#1#2{%
\def\includegraphic{}
\def\includegraphics{}
 \endgroup
 \patchcmd{\titleblock@produce}
  {\frontmatter@RRAPformat}
  {\frontmatter@RRAPformat{\produce@RRAP{*#1\href{mailto:#2}{#2}}}\frontmatter@RRAPformat}
  {}{}
}%
\makeatother
\begin{document}
\sethlcolor{yellow}
\preprint{AIP/123-QED}

\title[Characterizing current noise of commercial constant-current sources by using of an optically pumped rubidium atomic magnetometer]{Characterizing current noise of commercial constant-current sources by using of an optically-pumped rubidium atomic magnetometer}
\author{Ni Zhao}
 \affiliation{State Key Laboratory of Quantum Optics and Quantum Optics Devices and Institute of Opto-Electronics, Shanxi University, Taiyuan 030006,  Shanxi Province, China}
 
\author{Lulu Zhang}
 \affiliation{State Key Laboratory of Quantum Optics and Quantum Optics Devices and Institute of Opto-Electronics, Shanxi University, Taiyuan 030006,  Shanxi Province, China}

\author{Yongbiao Yang}
 \affiliation{State Key Laboratory of Quantum Optics and Quantum Optics Devices and Institute of Opto-Electronics, Shanxi University, Taiyuan 030006, Shanxi Province, China}

 \author{Jun He}
 \affiliation{State Key Laboratory of Quantum Optics and Quantum Optics Devices and Institute of Opto-Electronics, Shanxi University, Taiyuan 030006, Shanxi Province, China}
 \affiliation{Collaborative Innovation Center of Extreme Optics, Shanxi University, Taiyuan 030006, Shanxi Province, China}

 \author{Yanhua Wang}
  \affiliation{State Key Laboratory of Quantum Optics and Quantum Optics Devices and Institute of Opto-Electronics, Shanxi University, Taiyuan 030006, Shanxi Province, China}
 \affiliation{College of Physics and Electronic Engineering, Shanxi University, Taiyuan 030006, Shanxi Province, China}

 \author{Tingyu Li}
 \affiliation{College of Electronic Information and Optical Engineering, Taiyuan University of Technology, Taiyuan 030024, Shanxi Province, China}
 
\author{Junmin Wang}
\altaffiliation{corresponding author. E-mail: wwjjmm@sxu.edu.cn }
\affiliation{State Key Laboratory of Quantum Optics and Quantum Optics Devices and Institute of Opto-Electronics, Shanxi University, Taiyuan 030006, Shanxi Province, China}
\affiliation{Collaborative Innovation Center of Extreme Optics, Shanxi University, Taiyuan 030006, Shanxi Province, China}
\date{\today}

\begin{abstract}
This paper introduces a method for characterizing the current noise of  commercial constant-current sources(CCSs) using a free-induction-decay(FID) type optically-pumped rubidium atomic magnetometer driven by a radio-frequency(RF) magnetic field. 
We convert the sensitivity of the atomic magnetometer into the current noise of  CCS by calibrating the coil constant. At the same time, the current noise characteristics of six typical commercial low-noise CCSs are compared. The current noise level of the  KeySight Model B2961A is the lowest among the six tested CCSs, which is $36.233\pm0.022$  nA / Hz$^{1/2}$ at 1 $\sim$ 25 Hz  and  $133.905\pm0.080$ nA / Hz$^{1/2}$ at 1 $\sim$ 100 Hz respectively. The sensitivity of atomic magnetometer is dependent on the current noise level of the CCS. The CCS with low noise is of great significance for high-sensitivity atomic magnetometer. The research provides an important reference for promoting the development of high precision CCS, metrology and basic physics research. 
\end{abstract}

\maketitle
\section{INTRODUCTION}

Optically pumped atomic magnetometers mainly extract magnetic field information based on the interaction between light and atoms$^{1-3}$, which  have been widely used in military, medicine, space magnetic measurement, atomic gyroscope and basic physics research due to their outstanding advantages such as high sensitivity, fast response speed and portability$^{4-8}$.   According to their various working principles, optically pumped atomic magnetometers are mainly composed of the spin-exchange relaxation-free (SERF) atomic magnetometer$^{9}$, the nonlinear magneto-optical rotation(NMOR) atomic magnetometer$^{10}$, the coherent population trapping (CPT) atomic magnetometer$^{11}$, the Mx magnetometer$^{12}$, 
 the Mz magnetometer$^{13}$, etc.

Constant-current sources(CCSs) with low noise and excellent stability have important applications  in metrology, quantum precision measurement, search for neutron electric dipole moment$^{14-15}$ and basic physics research. 
 Current noise levels can be used to evaluate the characteristics of CCSs. Traditionally, the current noise of CCSs can be characterized indirectly based on Ohm's law. A constant current is applied to a high-precision resistance through a CCS. By analyzing the voltage signal noise on the resistance over a period of time, the current noise of the CCS can be deduced from Ohm's law.  However, the resistivity may be affected by thermal noise of resistance, which can affect the measurement results. Atoms are the most sensitive measurement media in nature. Based on the high sensitivity atomic magnetometer, the magnetic field can be accurately measured and the current noise of CCSs can be more precisely characterized.

 In recent years, significant progress has been made in characterizing and suppressing current noise. Shifrin $\emph{et al.}$$^{16}$realized high precision DC current measurement using a single-layer quartz two-zone solenoid and high precision differential current-frequency converter based on a He-Cs atomic magnetometer. Miao $\emph{et al.}$$^{17}$measured the frequency, amplitude and phase of sinusoidal alternating current using a pump-probe atomic magnetometer.  Li $\emph{et al.}$  $^{18}$developed a high-precision DC current sensor based on the optically pumped Mz atomic magnetometer.  Chen $\emph{et al.}$ $^{19}$characterized the current noise based on a pump-probe atomic magnetometer. Shen $\emph{et al.}$ $^{20}$ measured and suppressed the current noise of commercial CCS with a potassium atomic magnetometer.  Zheng $\emph{et al.}$ $^{21}$measured and suppressed the low-frequency noise of CCS based on double resonance alignment magnetometers.

The FID atomic magnetometer could operate in a large  range of terrestrial magnetic field, and have a relatively wide dynamic magnetic measurement range and high sensitivity$^{22-26}$. In our previous work$^{27}$, the  fundamental principle and classical physical picture were described in detail based on a FID atomic magnetometer driven by a RF  magnetic field.  Here, we present a method for characterizing current noise of commercial CCS by using the FID atomic magnetometer driven by a RF magnetic field. In this method, we calibrate coil constant in  magnetic field shields based on a high-precision commercial CCS. The measured magnetic field noise is converted to the current noise of commercial CCSs by the coil constant. We select six typical commercial CCSs (KeySight Model B2961A, ThorLabs Model LDC205C, SRS Model LDC501, SRS Model CS580, Home-made CCS and GWInstek Model 2303S)  to characterize and compare the current noise characteristics within the bandwidth range of 1 $\sim$ 25 Hz and 1 $\sim$ 100 Hz. The current noise characteristics of different commercial CCSs are analyzed and discussed in detail. We also experimentally demonstrate that the dependence between sensitivity and current noise.

\section{EXPERIMENTAL SETUP}
Figure 1 shows the experiment setup. A 15×15×15 mm$^{3}$ vapor cell containing isotopically enriched $^{87}$Rb is used in our experiment, which is filled with 100 Torr N$_{2}$ gas as  buffer gas and  fluorescence-quenching gas. The vapor cell is positioned at the center of the boron nitride ceramic oven. A specially designed square flexible film electric heater made by twisted pair wires is attached to the outer surface of the oven, which is used to heat and control the temperature of the atomic vapor cell. Here, the flexible film electric heater is driven by 477 kHz alternating current, which is set to be much higher than the measurement bandwidth and Larmor frequency to ensure that the heating system does not interfere with the measurement. And the non-magnetic PT100 thermistor is used as the temperature sensor without introducing magnetic interference.   A four-layer cylindrical µ-metal magnetic field shield is used to suppress environmental magnetic field noise.   The commercial CCSs apply current to produce a static magnetic field B$_{0}$ along the y direction and a RF magnetic field B$_{RF}$ along the z direction. The pump laser is emitted from a distributed Bragg reflector (DBR) laser, which is tuned to the $^{87}$Rb D1 transition line at 795 nm(from 5$^{2}$$\emph{S}_{1/2}$ $\emph{F}$=2 to 5$^{2}$$\emph{P}_{1/2}$ $\emph{F'}$=1). The pumped  beam passes through an acoustic-optical modulator(AOM), expands the beam through a telescope system, and is converted  into a circularly polarized beam through a $\lambda / 4$ wave plate, which enters the atomic vapor cell along the y direction. The diameter of the expanded beam is about 10 mm. The pump beam has a power of 5 mW. The linearly polarized probe laser, originated from a 780 nm DBR laser, is blue detuned by 6 GHz from the $^{87}$Rb D2 transition line at 780 nm(from 5$^{2}$$\emph{S}_{1/2}$ $\emph{F}$=1 to 5$^{2}$$\emph{P}_{3/2}$ $\emph{F'}$=2). The probe  beam has a diameter of 2 mm and a power of 30 $\mu$W. The direction of the probe  beam is perpendicular to the pump  beam and the RF magnetic field. The probe  beam passes through the atomic vapor cell and enters the polarimeter composed of $\lambda / 2$ wave plate, a Wollaston prism and a balanced differential photoelectric detector(common-mode noise rejection ratio $\sim$ 50 dB). We obtain information about the Faraday rotation angle by the data acquisition system composed of NI data acquisition card DAQ (NI-USB6363) and LabView.

The relationship betweent static magnetic field measured by the FID atomic magnetometer and the Larmor precession frequency can be expressed as:
\begin{flalign}
  \ B={\omega}/{\gamma}  
\end{flalign}

$\gamma$ represent the gyromagnetic ratio of ground state atoms, which is about 6.99583 Hz/nT of ground state (F = 2) of $^{87}$Rb.

\begin{figure}[h]
    \centering
    \includegraphics[width=0.53 \textwidth,height=0.4\textwidth]{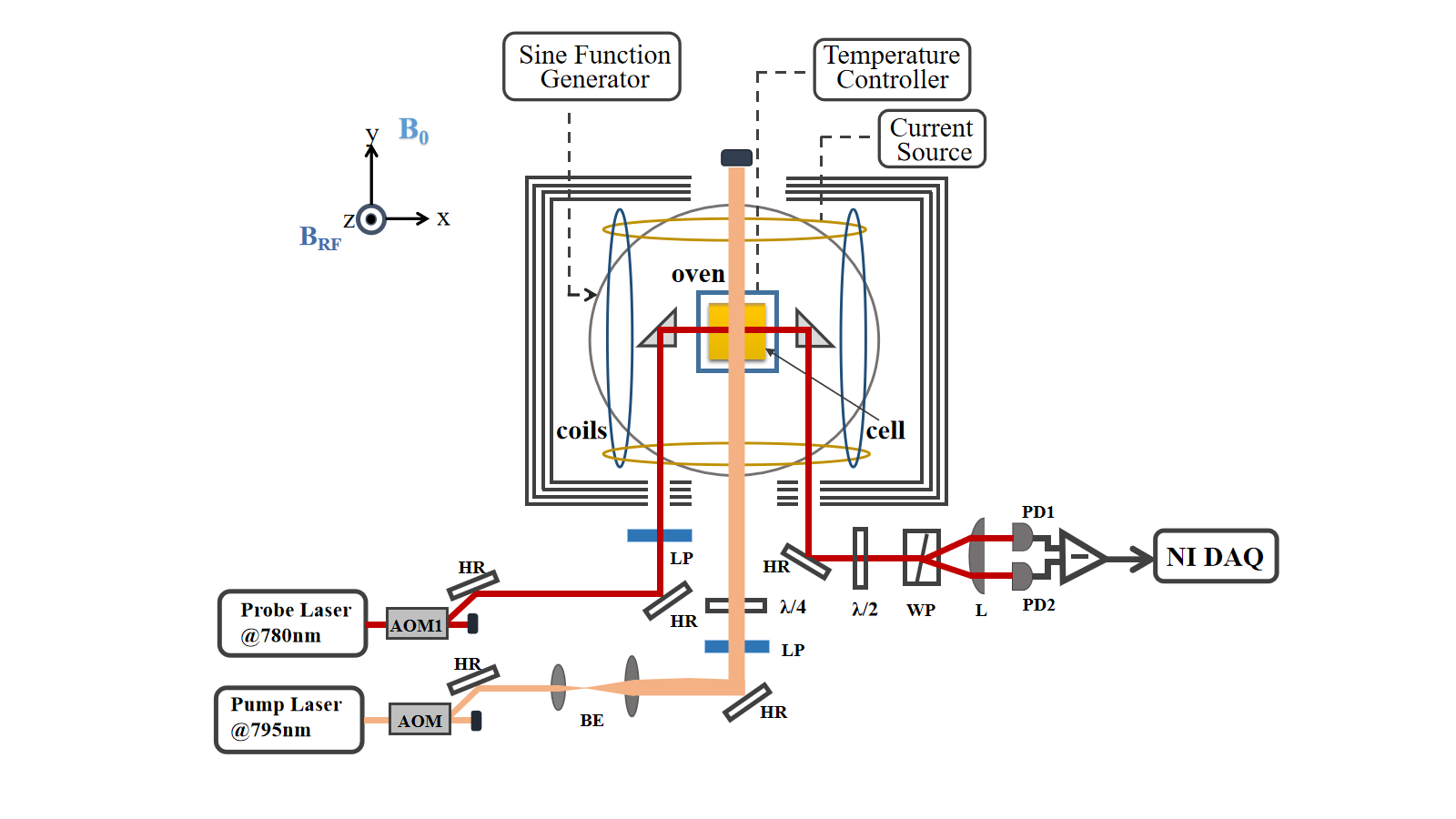}
    \caption{ Experimental setup of a FID type optically-pumped rubidium atomic magnetometer driven by a RF magnetic field. A 15×15×15 mm$^{3}$ vapor cell containing isotopically enriched $^{87}$Rb and 100 Torr N$_{2}$ gas is positioned at the center of a four-layer cylindrical µ-metal magnetic field shield. AOM: acousto-optical modulator; BE: beam expander; $\lambda / 4$ : quarter-wave plate;  $\lambda / 2$ : half-wave plate; LP: linear polarizer with a high extinction ratio; 
     WP: Wollaston prism;
     L:lens;
     PD1(PD2): balanced differential photoelectric detector; 
     NI DAQ: data acquisition.}
    \label{fig:my_label}
\end{figure}

The timing sequence diagram of control system is shown in Figure 2. At first, we switch on the pumped laser beam to prepare the spin-polarized state of the  $^{87}$Rb atomic ensemble from $t_0$ - $t_1$. The polarized  $^{87}$Rb atomic macroscopic magnetic moment is along the y direction at the end of $t_1$. Then, applying a RF magnetic field with the angular frequency equal to the Larmor precession frequency, the atomic macroscopic magnetic moment precess to the xoz plane after applying $\pi / 2$ pulse. Finally, the RF magnetic field is switched off and the probe laser beam is switched on. The atomic macroscopic magnetic moment evolve freely at Larmor frequency until the thermal equilibrium states. The pump laser , the RF magnetic field and the probe laser are separated from the time domain by time sequence control to avoid the crosstalk effect on the measurement signal and sensitivity, and further influence on the current noise characterization.  

We apply a static magnetic field of 6.3 $\mu$T along the y direction. The heating temperature of the atomic vapor cell is set at 85℃, and the atomic number density is about 2.2 × 10$^{12}$ cm$^{-3}$. Figure 3(a) shows a typical FID signal in one period. The  transverse relaxation times T$_2$ is 2.5 ms of $^{87}$Rb by exponential fitting. Figure 3(b) shows the Fast Fourier Transform (FFT) of FID signal. The full width at half maximum (FWHM) is $292.4\pm2.9$ Hz.

\begin{figure}[h]
    \centering
    \includegraphics[width=0.5 \textwidth,height=0.35\textwidth]{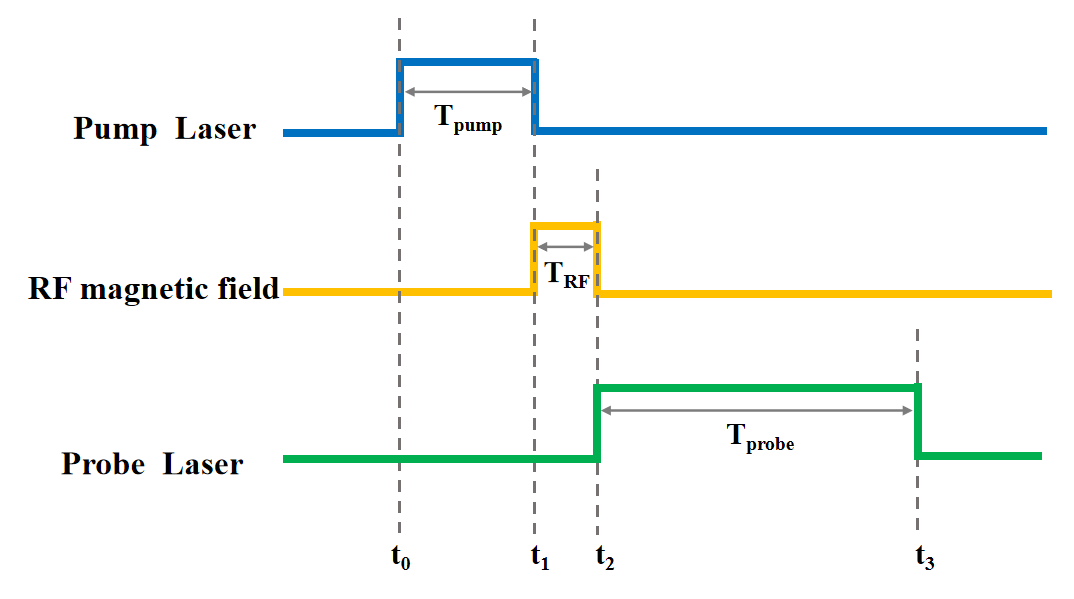}
    \caption{Time sequence control diagram for one period. The pump laser, the RF magnetic field and the probe laser are separated in time sequence. The pump laser beam is switched on during $t_0$-$t_1$; RF magnetic field is switched on for $\pi / 2$ 
    pulse during $t_1$-$t_2$; The probe laser beam is switched on during $t_2$-$t_3$.}
    \label{fig:my_label}
\end{figure}

\begin{figure}[h]
    \centering
    \includegraphics[width=0.5 \textwidth,height=0.5\textwidth]{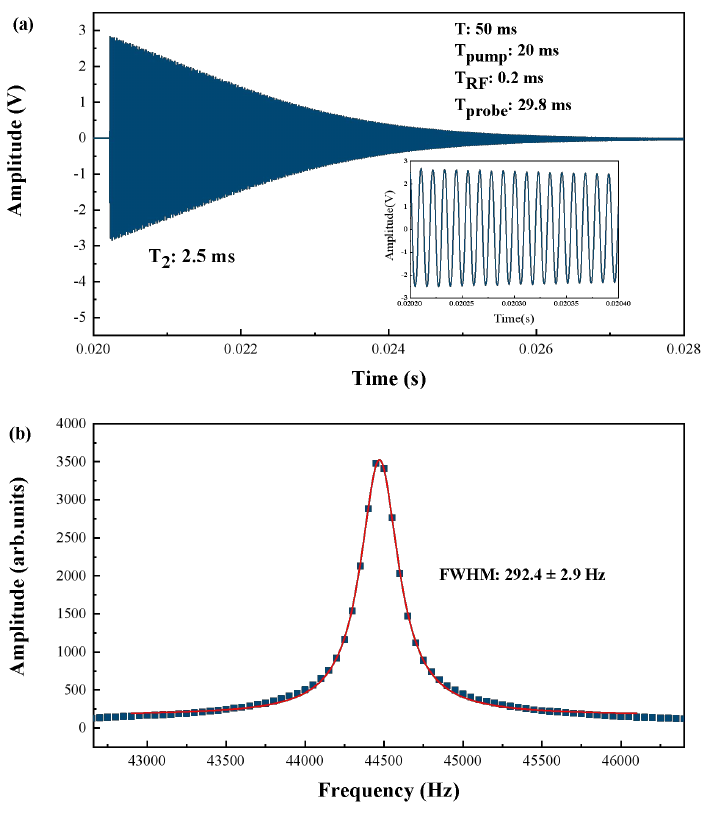}
    \caption{(a) FID signal with a period of 50 ms. The inset shows the zoomed-in view of the FID signal. The  transverse relaxation times T$_2$ is 2.5 ms of $^{87}$Rb by exponential fitting; (b) The FFT of the FID signal. The FWHM is $292.4\pm2.9$ Hz. }
    \label{fig:my_label}
\end{figure}

\section{MEASUREMENT RESULTS AND DISCUSSION}
\subsection{\label{sec:level3}Calibration of the coil constant}

In the experiment, a low-noise and high-stability  CCS KeySight Model B2961A  and a four-layer cylindrical µ-metal magnetic field shield provide a good conditions to calibrate the coil constant. The  coil constant can be described by$^{28-29}$: 

\begin{flalign}
  \ C_{coil}=B_{total}/{I}
\end{flalign}
where I is current, $B_{total}$ is the total magnetic field, which can be measured by using 
an atomic magnetometer.

Figure 4 shows result of the coil constant calibration along the y direction. At first, CCS B2961A apply a known current to the coils. Then we record the FID signal for 240 s when the period T is 50 ms. Larmor frequency is obtained by FFT transformation, and magnetic field value is obtained by calculation and statistical averaging. CCS B2961A applies current in the range of 2-250 mA. A series of magnetic field values are measured by using a FID magnetometer. The linear fitting results can be obtained as follows: 
\begin{flalign}
  \ B=126.956I-4.914  
\end{flalign}

The measured magnetic field is actually composed of the magnetic field generated by the CCS applying current to the coils and the residual magnetic field. According to the fitted linear equation(3), the calibrated coil constant is  $126.956\pm0.076$ nT/mA and the residual magnetic field is about 4.914 nT.
\begin{figure}[h]
    \centering
    \includegraphics[width=0.5 \textwidth,height=0.35\textwidth]{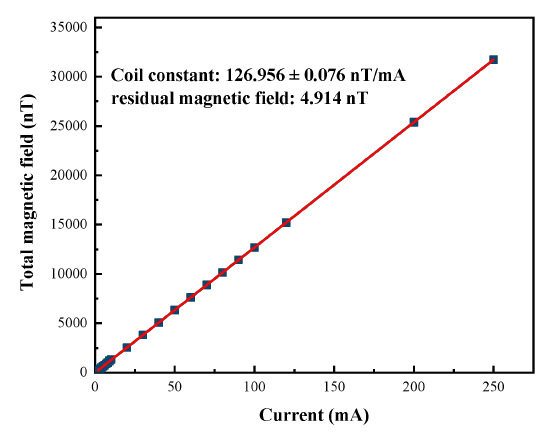}
    \caption{ Result of the coil constant calibration. CCS B2961A applies current from 2-250 mA to the coils along the y direction. The coil constant is  $126.956\pm0.076$ nT/mA and the residual magnetic field is about 4.914 nT.}
    \label{fig:my_label}
\end{figure}

\subsection{\label{sec:level3}Sensitivity analysis}

Sensitivity is an important index to evaluate the performance of atomic magnetometer. Taking B2961A as an example, we calculate and analyse the sensitivity. B2961A applies 100 mA current to the  coils along the y direction, and the static magnetic field is about 12.6 $\mu$T. We record 6000 periods FID signals by the data acquisition system and calculate sensitivity. Figure 5(a) shows the partially repeated measured FID signal with a sampling period of 5 ms. As shown in Figure 5(b), we obtained about 6000 DC magnetic field measurement values by converting the Larmor frequency into magnetic field values using Eq.(1). According to the statistical average of magnetic field values distribution, the static magnetic field is about 12.64154 $\mu$T. Figure 5(c) is the power spectral density(PSD) calculated with the magnetic field values, which shows a magnetic sensitivity is 17.0 pT/Hz$^{1/2}$ with a bandwidth of 1 $\sim$ 100 Hz. Here, we mainly measure and characterize the current noise of CCSs. Considering that the ambient magnetic field noise (for example, 1/ f noise, etc) is comparatively high at lower frequencies, in order to minimize the interference of various noises at low frequencies, we choose a bandwidth range of 1-100 Hz.

\begin{figure}[h]
    \centering
    \includegraphics[width=0.5 \textwidth,height=0.6\textwidth]{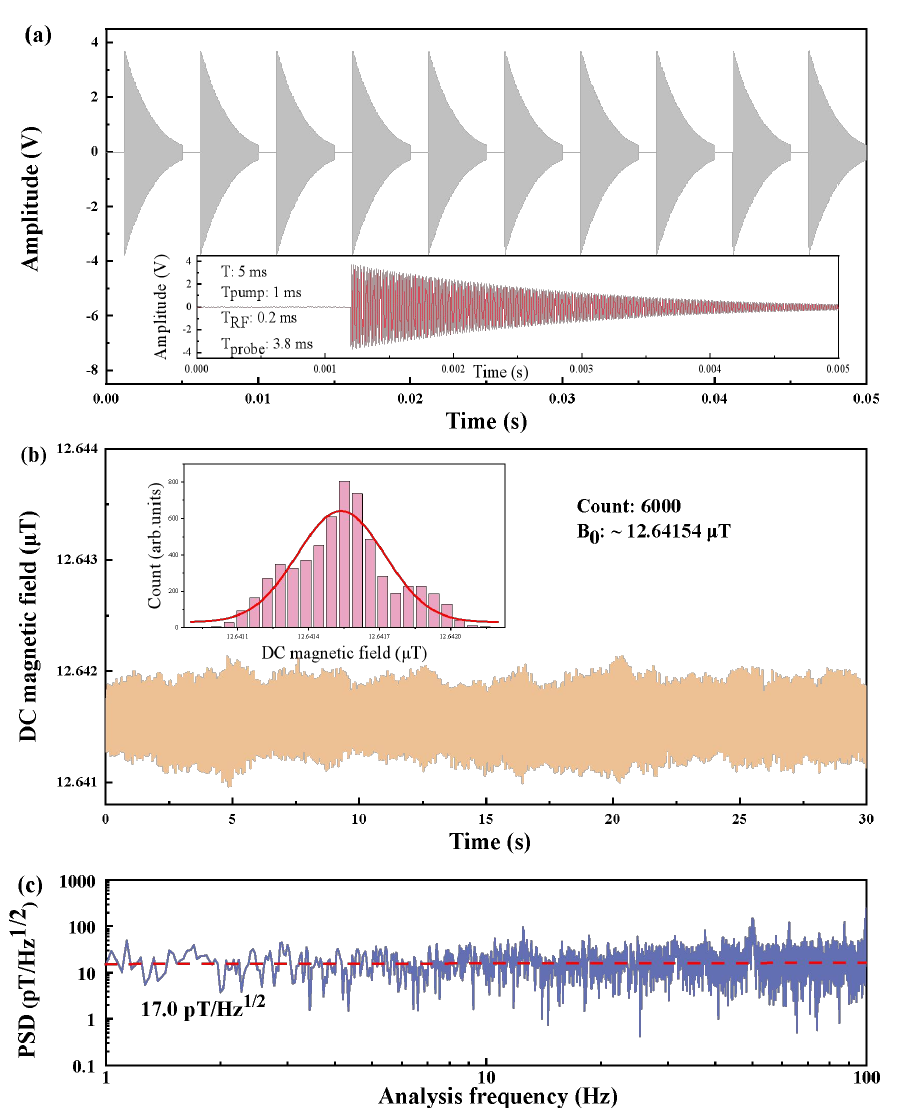}
    \caption{ (a) 0-0.05 s FID signal (inset: FID singal for one period); (b) Magnetic field values for 6000 sampling periods. The inset shows the statistical distribution of magnetic field values, and the mean value is approximately 12.64154 $\mu$T; (c) The PSD of magnetic field noise, which is about 17.0 pT/Hz$^{1/2}$ for 1 $\sim$ 100 Hz.}
    \label{fig:my_label}
\end{figure}

Figure 6  shows the electronic noise from the DAQ, the electronic noise from the photodetector with DAQ, and the intensity noise from the probe laser. The analysis shows that in the absence of
magnetic field, atomic ensemble, and other participation, the power spectral density of voltage noise obtained by using a photodetector
and DAQ (with the probe laser) is higher than the electronic noise of DAQ and photodetector (without the probe laser). In addition, DAQ has the lowest electronic noise. In other words, the electronic noise of DAQ and photodetector is not the main factor limiting the sensitivity. The sensitivity of the FID atomic magnetometer is mainly
limited by the transverse relaxation of the macroscopic spin magnetic moment of the atomic ensemble, the spin projection noise of the atomic ensemble, the intensity noise from the probe laser, and the current noise of the CCSs driving the magnetic field coils. The analysis of the sensitivity of different commercial CCSs makes plenty of sense.

\begin{figure}[h]
    \centering
    \includegraphics[width=0.5 \textwidth,height=0.35\textwidth]{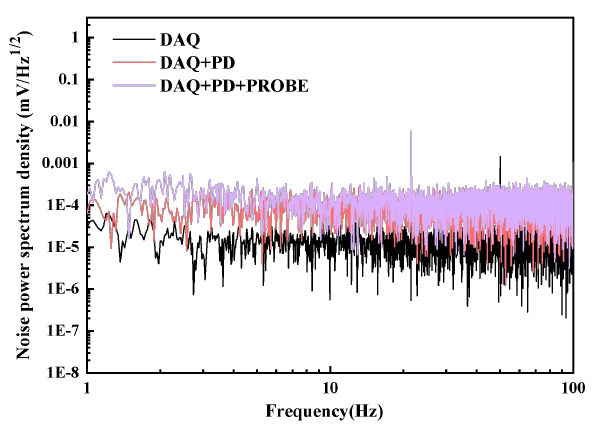}
    \caption{The noise floor of the system without atoms involved at 1-100Hz. The electronic noise from the DAQ(black line),  the electronic noise from the photodetector and the DAQ(pink line), and the intensity noise from the probe laser(purple line).}
    \label{fig:my_label}
\end{figure}

\subsection{\label{sec:level3}Characterization of current noise of commercial CCSs}

Six typical commercial CCSs (KeySight Model B2961A, ThorLabs Model LDC205C, SRS Model LDC501, SRS Model CS580, Home-made CCS and GWInstek 
 Model 2303S)  apply the same current of 100 mA (corresponding to the static magnetic field of 12.6 $\mu$T) to the  coils along the y direction. We obtain 6000 DC magnetic field values and analyze the sensitivity. Figure 7(a) and (b)  show the PSD for the atomic magnetometer with various CCSs for 1$\sim$ 25 Hz and for 1 $\sim$ 100 Hz respectively. The peak is caused by the 50-Hz electronic noise and its harmonic. The same experimental conclusions are presented for different CCSs that the sensitivity of the atomic magnetometer is improved significantly at 1 $\sim$ 25 Hz. Whether the frequency bandwidth is 1 $\sim$ 25 Hz or 1 $\sim$ 100 Hz, it can reflect the difference in the sensitivity of the magnetometer when the same current is generated by each CCS. The case using B2961A has the best sensitivity  among the six tested CCSs, which is 4.6 pT /Hz$^{1/2}$ for 1 $\sim$ 25 Hz and 17.0 pT/Hz$^{1/2}$ for 1 $\sim$ 100 Hz respectively.  The reason should be that CCS B2961A has ultra-low current noise and high-stability.

\begin{figure}
    \centering
\includegraphics[width=0.5\textwidth,height=0.5\textwidth]{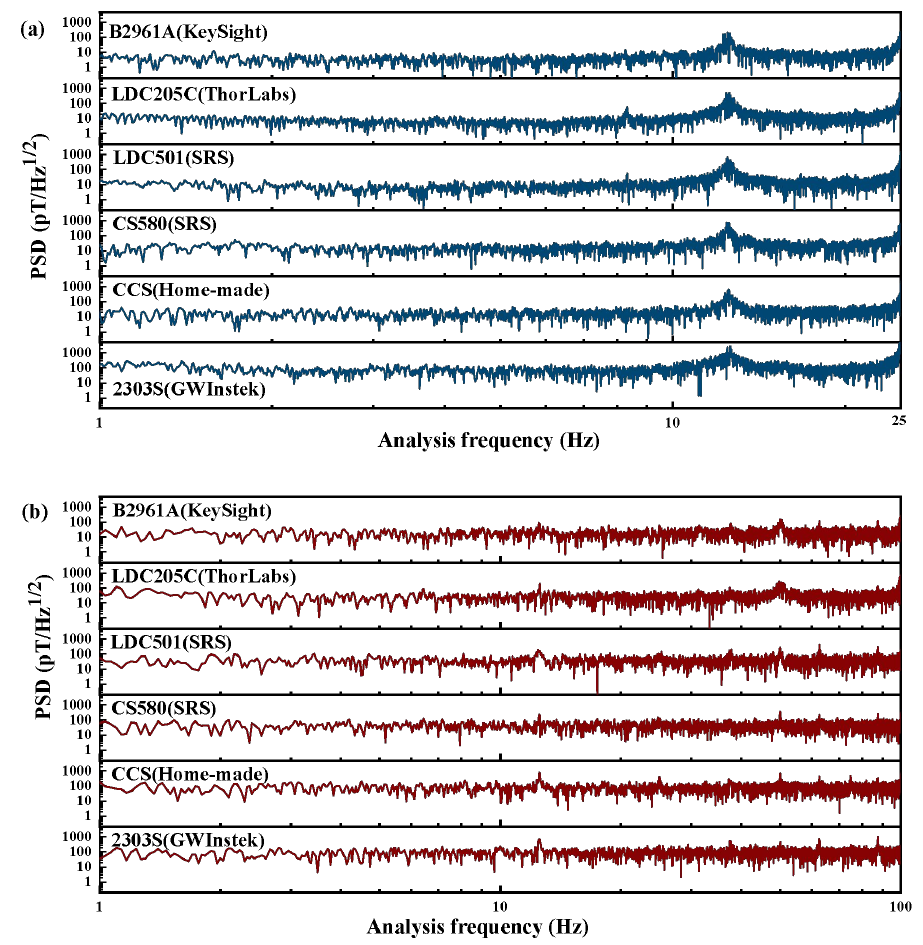}
    \caption{The PSD for the atomic magnetometer with various CCSs (labeled in figure) and two typical analysis freuqnecy ranges (1 $\sim$ 25 Hz, (a) and 1 $\sim$ 100 Hz, (b)).}
    \label{fig:my_label}
\end{figure}

In our experiment, the magnetic field of the atomic manetometer is generated by the CCSs, so the current noise of CCSs can be reflected from magnetic field noise power spectrum density. The sensitivity and current noise of different commercial CCSs are shown in Table 1. The current noise is obtained by dividing the sensitivity by the coil constant. The sensitivity of FID magnetometer is 17.0 pT/Hz$^{1/2}$ for 1 $ \sim $ 100 Hz. Dividing this value by the coil constant, the current noise is 133.905 $\pm$ 0.080 nA/Hz$^{1/2}$ when the CCS B2961A outputs a current of 100 mA. The sensitivity is 4.6 pT/Hz$^{1/2}$ at 1 $\sim$ 25 Hz, and the current noise is 36.233 $\pm$ 0.022 nT/mA  when the CCS B2961A outputs a current of 100 mA. It can also be clearly seen from the figure that different commercial CCSs have different current noise levels. The KeySight Model B2961A has the lowest current noise level among the six tested CCSs. ThorLabs Model LDC205C and SRS Model LDC501 have similar current noise levels. SRS Model CS580 and Home-made CCS have higher current noise levels. The GWInstek Model 2303S has the highest current noise level among the six tested CCSs. CCSs have lower current noise levels for 1 $\sim$ 25 Hz. The current noise level can clearly reflect the output current fluctuation of CCSs.

\begin{table*}
    \renewcommand{\arraystretch}{1.5}
 \caption{The sensitivity and current noise of different commercial CCSs.}
    \label{table_example}
    \centering
    \begin{tabular}{|c|c|c|c|c|}
        \hline
    {} & \multicolumn{2}{|c|} {Sensitivity(pT/Hz$^{1/2}$)} & 
        \multicolumn{2}{c|}{Current noise(nA/Hz$^{1/2}$)}\\
        \cline{2-5}
          &  Bandwidth(1 $\sim$ 25 Hz)&  Bandwidth(1 $\sim$ 100 Hz) &  Bandwidth(1 $\sim$ 25 Hz)   &  Bandwidth (1 $\sim$ 100  Hz)\\
        \hline
         B2961A(KeySight)    &   4.6 &  17.0 &   $36.233\pm0.022$ & $133.905\pm0.080$ \\
         \hline
         LDC205C(ThorLabs)    &   8.3 &  27.2&   $65.377\pm0.039$ &  $214.247\pm0.128$ \\
         \hline
        LDC501(SRS)  &  9.8 &  34.6 &  $77.192\pm0.046$&  $272.535\pm0.163$\\
         \hline
       CS580(SRS)    &   17.9 &  41.0 &   $140.994\pm0.084$ &  $322.947\pm0.193$\\
         \hline
      CCS(Home-made)   &   19.1 &  69.8 &   $150.446\pm0.090$ &$549.797\pm0.329$ \\
         \hline
       2303S(GWInstek)   &   73.5 &  87.9 &   $578.941\pm0.347$ &  $692.366\pm0.414$ \\
        \hline
    \end{tabular}
\end{table*}

\section{DISCUSSION AND CONCLUSION}
We present a method to characterize  current noise of different commercial CCSs based on the calibration of coil constant by using a FID atomic magnetometer. The sensitivity of atomic magnetometer is interdependent with the current noise of CCSs. The current noise of CCSs can be estimated by the sensitivity of atomic magnetometer. We characterize and compare the current noise characteristics within the analysis frequency range of 1 $\sim$
25 Hz and 1 $\sim$ 100 Hz. Bandwidth and sensitivity are mutually restricted, increasing bandwidth yields that sensitivity is getting worse. The sampling period of FID signal is longer, the Larmor frequency obtained after FFT is more accurate, the smaller the magnetic field fluctuation value over a period of time and the better the sensitivity obtained by calculation when a small bandwidth range is selected. As a result,  we characterize current noise more accurately. The CCS with low-noise and high-stability is of great significance to improve the sensitivity of atomic magnetometer. 

In addition, the sensitivity of an optically pumped magnetometer is limited by various factors, such as: the photon shot noise(PSN), the spin-projection noise(SPN), the fluctuations of the residual magnetic field, the intensity noise from probe laser, the electronic noise, etc. The various noise mentioned above are included in the collected FID signal, as well as in the calculated PSD and the measurement results of the current noise. Therefore, our measurement results are actually the upper bound of the current noise of  CCSs. It is also very important to further improve the sensitivity of magnetometer.  The spin-exchange  collisions(SE) among alkali-metal atoms has a significant effect on transverse spin-relaxation rates and the linewidth of magnetic resonance spectrum$^{30}$, which leads to a decrease in the sensitivity of atomic magnetometers. SE can be suppressed by filling buffer gas with appropriate pressure. Spin-destruction collisions can be suppressed by filling buffer gas or coating the atomic vapor cell's inner wall with anti-relaxation film. PSN and SPN can be suppressed by squeezed states of light field $^{31-32}$and atom spin squeezing$^{33}$.

 In the future, we can improve the sensitivity based on the following two methods: (\textrm{i})We can perform active magnetic field stabilization$^{34}$ based on CCSs with high stability and low noise, which can be further compensate and shield ambient magnetic field noise, and (\textrm{ii}) we can also introduce the polarization-squeezed light to further suppress  PSN, so that the sensitivity of the atomic magnetometer can go beyond PSN limit and realize  quantum enhancement measurement of the atomic magnetometer. 

\section*{ACKNOWLEDGMENTS}
This research was financially supported by the National Natural Science Foundation of China (11974226). 
\section*{AUTHOR DECLARATIONS}
\subsection*{Conflict of Interest}
The authors have no conflicts to disclose.
\subsection*{Author Contributions}
$\textbf{ Ni Zhao:}$ Data curation (equal); Formal analysis(equal); Writing–original draft(equal). {\bf Lulu Zhang:} Data curation (equal); Formal analysis(equal); Software (equal). {\bf Yongbiao Yang:} Data curation (equal). {\bf Jun He:} Investigation (supporting). {\bf Yanhua Wang:} Investigation (supporting); Funding acquisition(equal). {\bf Tingyu Li:} Investigation (supporting); Funding acquisition(equal). {\bf Junmin Wang:} Project administration (leading); Resources (leading); Funding acquisition (leading); Writing–review\& editing (leading); Supervision (leading).
\section*{DATA AVAILABILITY} 
The data that support the findings of this study are available from the corresponding author upon reasonable request.
\section*{References} 
$^1$ F. Bitter, “The optical detection of radiofrequency resonance,” Phys. Rev. {\bf 76,}    833-835 (1949).\\
$^2$ W. E. Bell and A. L. Bloom, “Optical detection of magnetic resonance in alkali metal vapor,” Phys. Rev. {\bf 107,} 1559-1565 (1957).\\
$^3$ D. Budker and M. Romalis, “Optical magnetometry,” Nature Phys. {\bf 3,} 227-234 (2007).\\
$^4$ M. E. Limes, E. L. Foley, T. W. Kornack, S. Caliga, S. McBride, A. Braun, W. Lee, V. G. Lucivero, and M. V. Romalis, “Portable magnetometry for detection of biomagnetism in ambient environments,” Phys. Rev. Applied {\bf 14,} 011002 (2020).\\
$^5$ T. H. Sander, J. Preusser, R. Mhaskar, J. Kitching, L. Trahms, and S. Knappe, “Magnetoencephalography with a chip-scale atomic magnetometer,” Biomed. Opt. Express {\bf 3,} 981–990 (2012). \\
$^6$ O. Alem, T. H. Sander, R. Mhaskar, J. LeBlanc, H. Eswaran, U. Steinhoff, Y. Okada, J. Kitching, L. Trahms, and S. Knappe, “Fetal magnetocardiography measurements with an array of microfabricated optically pumped magnetometers,”   Phys. Med. Biol. {\bf 60,} 4797–4811 (2015).\\
$^7$  H. Korth, K. Strohbehn, F. Tejada, A. G. Andreou, J. Kitching, S. Knappe, S. J. Lehtonen,  S. M. London, and M. Kafel, “Miniature
atomic scalar magnetometer for space based on the rubidium isotope $^8$$^7$Rb,” J. Geophys. Res. Space Physics {\bf 121,} 7870–7880 (2016).\\
$^8$ R. J. Li, W. Quan, W. F. Fan, L. Xing, and J. C. Fang, “Influence of magnetic fields on the bias stability of atomic gyroscope
operated in spin-exchange relaxation-free regime,” Sensors and Actuators A {\bf 266,} 130–134 (2017). \\
$^9$ J. C. Allred, R. N. Lyman, T. W. Kornack, and M. V. Romalis, “High-sensitivity atomic magnetometer unaffected by spin-exchange relaxation,”  Phys. Rev. Lett. {\bf 89,} 130801 (2002). \\
$^1$$^0$ W. Gawlik, L. Krzemień, S. Pustelny, D. Sangla, and J. Zachorowski, “Nonlinear magneto-optical rotation with amplitude modulated light,” Appl. Phys. Lett. {\bf 88,} 131108 (2006).\\
$^1$$^1$J. Belfi, G. Bevilacqua, V. Biancalana, S. Cartaleva, Y. Dancheva, and L. Moi, “Cesium coherent population trapping magnetometer for cardiosignal detection in an unshielded environment,” J. Opt. Soc. Am. B {\bf 24,} 2357–2362 (2007).\\
$^1$$^2$ S. R. Su, G. Y. Zhang, X. Bi, X. He, W. Q.  Zheng, and Q. Lin, “Elliptically polarized laser-pumped Mx magnetometer towards applications at room temperature,” Opt. Express  {\bf 27,} 33027-33039 (2019).\\
$^1$$^3$ Y. Gu, R. Y. Shi, and Y. H. Wang, “Study on sensitivity-related parameters of distributed feedback laser-pumped cesium atomic magnetometer,” Acta Physica Sinica {\bf 63,} 110701 (2014).(in Chinese) \\
$^1$$^4$ M. Jiang, H. W. Su, A. Garcon, X. H. Peng, and D. Budker, “Search for axion-like dark matter with spin-based amplifiers,” Nature. Phys. {\bf 17,} 1402–1407(2021).\\
$^1$$^5$ J. M. Pendlebury, K. F. Smith, R. Golub, J. Byrne, T. J. L. McComb, T. J. Sumner,
S. M. Burnett, A. R. Taylor, B. Heckel, N. F. Ramsey, K. Green, J. Morse, A. I. Kilvington, C. A. Baker, S. A. Clark, W. Mampe, P. Ageron, and P. C. Miranda, “Search for a neutron electric dipole moment,” Phys. Lett. B {\bf 136,} 327–330 (1984) .\\
$^1$$^6$ V. Y. Shifrin, V. N. Khorev, and P. G. Park, “A high-precision system for direct  current reproduction based on atomic magnetic resonance in helium-4,” Metrologia {\bf 36,} 171-177 (1999).\\
$^1$$^7$ P. X. Miao, “Measurement of 10 kHz sinusoidal alternating current by pump-probe atomic magnetometer,” Vacuum and Cryogenics  {\bf 28,} 592-600 (2022). (in Chinese)\\
$^1$$^8$ G. Z. Li, Q. Xin, X. X. Geng, Z. Liang, S. Q. Liang, G. M. Huang, G. X. Li, and G. Q. Yang, “Current sensor based on an atomic magnetometer for DC application,” Chinese Opt. Lett. {\bf 18,} 031202 (2020).\\
$^1$$^9$ D. Y. Chen, P. X. Miao, Y. C. Shi, J. Z. Cui, Z. D. Liu, J. Chen, and K. Wang, “Measurement of noise of current source by pump-probe atomic magnetometer,” Acta Physica Sinica  {\bf 71,} 024202 (2022). (in Chinese)\\
$^2$$^0$ L. Shen, R. Zhang, T. Wu, X. Peng, S. Yu, J. B. Chen, and H. Guo, “Suppression of current source noise with an atomic magnetometer,” Rev. Sci. Instrum. {\bf 91,} 084701 (2020).\\
$^2$$^1$ J. T. Zheng, Y. Zhang, Z. Y. Yu, Z. Q. Xiong, H. Luo, and Z. G. Wang, “Precision measurement and suppression of low-frequency noise in a current source with double-resonance alignment magnetometers,” Chinese Phys. B {\bf 32,} 040601(2023).\\
$^2$$^2$ R. Zhang, D. Kanta, A. Wickenbrock, H. Guo, and D. Budker, “Heading-error-free optical atomic magnetometry in the earth-field range,” Phys. Rev. Lett. {\bf 130,} 153601 (2023).\\
$^2$$^3$ S. R. Li, D. Y. Ma, K. Wang, Y. N. Gao, B. Z. Xing, X. J. Fang, B. C. Han, and W. Quan, “High sensitivity closed-loop Rb optically pumped magnetometer for measuring nuclear magnetization” 
 Opt. Express   {\bf 30,} 43925-43937 (2022).\\
$^2$$^4$ D. Sheng, S. Li, N. Dural, and M. V. Romalis, “Subfemtotesla scalar atomic magnetometry using multipass cells,”  Phys. Rev. Lett. {\bf 110,} 160802 (2013).\\
$^2$$^5$ D. Hunter, S. Piccolomo, J. D. Pritchard, N. L. Brockie,T. E. Dyer, and E. Riis, “Free-induction-decay magnetometer based on a microfabricated Cs vapor cell,”  Phys. Rev. Applied. {\bf 10,} 014002(2018).\\
$^2$$^6$ C. Liu, H. F. Dong, and J. J. Sang, “Submillimeter-resolution magnetic field imaging with digital micromirror device and atomic vapor cell,” Appl. Phys. Lett. {\bf 119,} 114002 (2021).\\
$^2$$^7$ L. L. Zhang, Y. B. Yang, N. Zhao, J. He, and J. M. Wang, “A multi-pass optically pumped rubidium atomic magnetometer with free induction decay” Sensors {\bf 22,} 7598 (2022).\\
$^2$$^8$ L. L. Chen, B. Q. Zhou, G. Q. Lei, W.F. Wu, J. Wang, Y. Y. Zhai, Z. Wang, and J. C. Fang, “A method for calibrating coil constants by using the free induction decay of noble gases” AIP Advances  {\bf 7,} 075315 (2017). \\
$^2$$^9$ H. Zhang, S. Zou, and X. Y. Chen, “A method for calibrating coil constants by using an atomic spin co-magnetometer,” Eur. Phys. J. D . {\bf 70,} 203(2016).\\
$^3$$^0$ W. Happer, “Optical Pumping,” Rev. Mod. Phys. {\bf 44,} 169(1972). \\
$^3$$^1$ C. Troullinou, R. Jiménez-Martínez, J. Kong,  V. G. Lucivero, and M. W. Mitchell, “Squeezed-light enhancement and backaction evasion in a high sensitivity optically pumped magnetometer,” Phys. Rev. Lett. {\bf 127}, 193601(2021).\\
$^3$$^2$ L. L. Bai, X. Wen, Y. L. Yang, L. L.Zhang, J. He,  Y. H. Wang, and J. M. Wang, “Quantum-enhanced rubidium atomic magnetometer
based Faraday rotation via 795-nm  Stokes operator squeezed light,” J. Opt. {\bf 23}, 085202(2021).\\
$^3$$^3$ H. Bao, J. L. Duan, S. C. Jin, X. D. Lu, P. X. Li, W. Z. Qu,  M. F. Wang, I. Novikova,  E. E. Mikhailov,  K. F. Zhao, K. Mølmer, H. Shen, and Y. H. Xiao, “Spin squeezing of 10$^{11}$
atoms by prediction and retrodiction measurements,” Nature   {\bf 581}:159–163(2020).\\
$^3$$^4$ Y. D. Ding, R. Zhang, J. H. Zheng, J. B.  Chen, X. Peng, T. Wu, and H. Guo, “Active stabilization of terrestrial magnetic field with potassium atomic magnetometer,” Rev. Sci. Instrum. {\bf 93,} 015003(2022). \\

\end{document}